\documentclass{article}
\input amssym.def
\input amssym.tex
\usepackage[a4paper]{geometry}

\def\ben{\begin{equation}}
\def\een{\end{equation}}
\def\half{\frac{1}{2}}
\def\bea{\begin{eqnarray}}
\def\eea{\end{eqnarray}}

\def\cR{{\cal R}}

\def\bx{{\bf x}}

\def\bE{{\bf E}}\def\bB{{\bf B}}
\def\p{\partial}
\def\mathbb{\Bbb}
\def\cL{{\cal L}}
\def\ben{\begin{equation}}
\def\een{\end{equation}}
\def\half{{1 \over 2}}
\def\bea{\begin{eqnarray}}
\def\eea{\end{eqnarray}}

\def\nn{\nonumber}

\def \p{\partial}

\def\p{\partial}
\def\babla{{\mbox{\boldmath $ \nabla $ }}} 
 
\def\cL{{\cal L}}

\def\Z{{\Bbb  Z}}\def\R{{\Bbb  R}}
\def\E{{\Bbb  E}}

\newcount\hour \newcount\minute
\hour=\time  \divide \hour by 60
\minute=\time
\loop \ifnum \minute > 59 \advance \minute by -60 \repeat
\def\nowtwelve{\ifnum \hour<13 \number\hour:
                      \ifnum \minute<10 0\fi
                      \number\minute
                      \ifnum \hour<12 \ A.M.\else \ P.M.\fi
         \else \advance \hour by -12 \number\hour:
                      \ifnum \minute<10 0\fi
                      \number\minute \ P.M.\fi}
\def\nowtwentyfour{\ifnum \hour<10 0\fi
                \number\hour:
                \ifnum \minute<10 0\fi
                \number\minute}

\begin{document}

\title{
\begin{flushright}
 \small{UPR-1235-T \\ DAMTP-2012-2}
\end{flushright}
\vspace{0.2in}
 { \Large Conformal Symmetry of a  Black Hole as  a Scaling Limit:\\
 A Black Hole in an Asymptotically Conical Box
 }}
\author{\large M. Cveti\v c$^{1,2}$ and G.W Gibbons $^{3}$  
\\
\\   \small{$^1${\it Department of Physics and Astronomy,
University of Pennsylvania, Philadelphia, PA 19104-6396, USA} }
\\ \small{$^2${\it  Center for Applied Mathematics and Theoretical Physics,
University of Maribor, Maribor, Slovenia}}
\\   \small{$^3${\it  D.A.M.T.P., University of  Cambridge, 
Wilberforce Road, Cambridge CB3 0WA, U.K. }              }  
\\ }
\maketitle
\begin{abstract}
We show that the  previously  obtained 
subtracted geometry of  four-dimensional asymptotically flat multi-charged  
rotating black holes,
whose massless wave equation  exhibit $SL(2,\R) \times SL(2,\R) \times SO(3)$ symmetry
may  be obtained by a suitable 
scaling limit of certain asymptotically 
flat multi-charged rotating black holes,
which is reminiscent of   
 near-extreme black holes in the dilute gas approximation.  
The co-homogeneity-two geometry is supported by a dilation field and two 
 gauge-field strengths. We also point out  that these  subtracted
geometries can be obtained as a particular  Harrison transformation of 
the original  black holes. 
Furthermore the  subtracted metrics are  asymptotically 
conical (AC),   like   global monopoles, thus  describing  ``a black hole in an AC box''.
Finally  we account for the  the emergence of   
the $SL(2,\R) \times SL(2,\R) \times SO(3)$ symmetry as a 
consequence  of the subtracted metrics
being  Kaluza-Klein type  quotients of   $ AdS_3\times 4  S^3$.  We demonstrate that similar properties hold for five-dimensional black holes.
\end{abstract}
\eject
\pagebreak

\tableofcontents
\addtocontents{toc}{\protect\setcounter{tocdepth}{2}}

\newpage

\section{Introduction}

The microscopic  entropy of  (near)-supersymmetric 
asymptotically flat black holes has  been well understood  
in terms of  weakly coupled two-dimensional conformal field theory 
(See, e.g.,  the review \cite{Sen} and references therein).  
On the other hand,   general multi-charged rotating  black holes  
in four \cite{CYI} and five  \cite{CYII} dimensions  have  
an  entropy formula \cite{CYI} strongly suggestive  of a possible 
microscopic interpretation in terms of a 
 weakly coupled two-dimensional conformal field theory. 
Some early work along these directions was 
pursued in \cite{Larsen,CL97I,CL97II}.  [There are indications that general asymptotically anti-deSitter black holes may  also have  a microscopic description, as indicated by the quantized value of the product of  horizon areas \cite{CGP}.]

An intriguing  clue to the internal structure of a black hole is 
the structure of  the wave equation  in its background.   
The wave equation for a massless scalar field turns out to have  
remarkable simplifications 
even for general multi-charged rotating black holes \cite{CL97I, CL97II} .  
In particular the equation is separable and it has  an 
$SL(2, \R)\times SL(2,\R) \times SO(3)$
symmetry, when certain terms are subtracted. 
It turns out that these terms can be neglected in many special cases, 
including the near-supersymmetric limit (the AdS/CFT correspondence) 
\cite{MS,CL97I,CL97II}, 
the near extreme rotating limit (the Kerr/CFT correspondence) 
\cite{GuicaMU,CL09}, and the low energy limit \cite{CL97I,CastroFD}.
 However, in general there is no $SL(2, \R)\times SL(2,\R) \times SO(3)$
isometry of the subtracted metrics. In  \cite{CastroFD}   
it is asserted that the $SL(2, \R)\times SL(2,\R) \times SO(3)$  symmetry 
suggested by the massless wave equation is there--it is just that it is 
spontaneously broken (``hidden conformal symmetry'').

 Recently, in \cite{CL11I}  an explicit part of the  general multi-charged rotating black hole geometry, which exhibits the $SL(2,\R)\times SL(2,\R) \times SO(3)$ symmetry of the  the wave equation, was constructed.  
Since  this metric differs from  the original
 black hole metric by  removing certain terms in the warp factor, only,   
it was dubbed  the ``subtracted geometry''.   
The subtracted metric has the same  horizon area  and 
periodicity's of the angular and time 
coordinates in the near horizon regions. 
It is thus expected to preserve the internal structure of the black hole.
The subtracted  geometry is not asymptotically flat (AF) 
but is asymptotically conical, AC,  and admits a Lifshitz type
homothety  which scales space  and time differently.   
 The physical interpretation of this subtraction is 
the removal of the  ambient asymptotically Minkowski space-time in a 
way that extracts the  $SL(2,\R)\times SL(2,\R) \times SO(3)$ 
symmetry of the black hole (``black hole in an AC confining  box'').

 The subtraction has been explicitly implemented both for the 
five-dimensional three-charge rotating black holes \cite{CL11I}
 and four-dimensional four-charge ones \cite{CL11II}.   
For four-dimensional black holes the subtracted  geometry is a Kaluza-Klein 
type quotient       of $AdS_3\times 4 S^2$  with   
$SL(2,\R)\times SL(2,\R) \times SO(3)$ symmetry  manifest.  
 [Analogously, for five-dimensional black holes the subtracted geometry is as Kaluza-Klein type quotient of $AdS_3\times S^3$ with  $SL(2,\R)\times SL(2,\R) \times SO(4)$ symmetry
 manifest.]
In \cite{CL11II}  the explicit sources for static 
subtracted geometry were obtained as a part of the so-called 
STU-model (sector of the 
four-dimensional {\cal N}=2 supergravity coupled to three vector 
supermultiplets) and  its lift to five dimensions  corresponds to  
the minimal supergravity.
 
In this paper  we further elucidate  the origin and the 
geometric interpretation of the subtracted geometries. 
In particular we realize these geometries for general four-charge rotating solutions as a scaling limit of 
certain multi-charge rotating black holes with three large charges, 
reminiscent of the near-extreme multi-charge rotating
black holes in the dilute gas approximation.
This procedure fully determines  the sources  
in the rotating multi-charge cases both in four and five dimensions.

We  also demonstrate that  the subtracted geometry of the Schwarzschild solution can be obtained by performing a
 Harrison transformation on the original  Schwarzschild solution. This procedure confirms tat  the  subtracted geometry
 is a solution of the STU-model. 
 In the previous version of this paper we conjectured that the subtracted geometry of general multi-charged rotating black holes  arises as a  Harrison transformation of the original 
 multi-charged rotating black hole. This has since been confirmed \cite{Vir}.
These results  show that the original black hole and  the subtracted geometry clearly lie in the same duality orbit, specified by a Harrison transformation  and  passing through the original black hole. 
Thus any physical property of the original black hole  solution which is invariant under duality transformation of the theory  remains a property of the subtracted geometry.

Furthermore we analyze  the asymptotic structure of 
the subtracted  metrics and find,  because the energy density falls
of inversely as the second power of the radial   distance
that they are  asymptotically
conical (AC), rather than asymptotically flat,  
in a way which is similar to the asymptotic  behaviour of
global monopoles and isothermal gas spheres. Since 
the metric component $g_{tt}$ is proportional to the $6$th
power of the radial area distance, the subtracted metrics exhibit 
confining behaviour analogous to that found in $AdS_4$, justifying
their  interpretation  as ``a  black hole in  an AC  confining box''.
The spatial dependence of   the dilaton implies that the gauge coupling
constants run  logarithmically in the radial direction, not even stabilizing at infinity. 
This  behaviour, together with energy densities falling off 
inversely as the square of the radial distance 
is shown to persist  in Dilaton-Maxwell theory when  a  limit        
of vanishing Newton's constant is taken.

The paper is organized in the following way: In Section 2  we  obtain the subtracted geometry of general four-charge rotating black holes  (of ${\cal N}=2$ supergravity coupled to three vector supermultiplets)   as a scaling limit of black holes  holes with three large charges, accounting for 
all the sources  both in the static  (Section 2.1) and  rotating (Section 2.2) case. In Section 2.3 we demonstrate for the Schwarzschild solution that the subtracted geometry emerges as a c Harrison transformation on the unsubtracted one. 
In Section 3 we further discuss asymptotics  of the subtracted geometry and draw a comparison with other  AC examples in General Relativity, in particular those of isothermal gas spheres and  a global monopole (Section 3.1). The confining properties of the AC metrics are discussed in Section 3.2.  In Section 4 we  discuss the local and global properties of the geometry which is a Kaluze-Klein coset of $AdS_2\times 4 S^2$ (Section 4.1), derive the form of the isometry generators for the geometry lifted to five-dimensions (Section 4.2)  and discuss isometries of the five-dimensional metric constrained to special slices (Section 4.3).   Conclusions are given in Section 5. In Appendix we present the full solution of the subtracted geometry for general three-charge black holes in five dimensions.
\section{Subtracted Geometry as a Scaling Limit}

In this section we obtain the explicit expressions for  sources that support the subtracted geometry of general four-charge rotating black holes by taking a scaling limit of a rotating black
hole  with three equal large charges and the fourth  finite one. The limit is closely related to the ``dilute gas'' (near-BPS) rotating black hole solution.  

The original four-charge rotating solution \cite{CYI}, along with the  explicit expressions for all four gauge potentials was given
 in \cite{CCLPI} as a solution of the  bosonic sector four-dimensional Lagrangian density  of the
${\cal N}=2$ supergravity coupled to three vector supermultiplets
 \footnote{Conventions for dualisation in \cite{CCLPI} are that a
$p$-form $\omega$ with components defined by $\omega=1/p!\,
\omega_{i_1\cdots i_p}\, dx^{i_1}\wedge \cdots \wedge dx^{i_p}$ has
dual ${*\omega}$ with components $({*\omega})_{i_i\cdots i_{D-p}} =
1/p!\, \epsilon_{i_1\cdots i_{D-p} j_i\cdots j_p}\, \omega^{j_1\cdots
j_p}$. Note also that  the normalization of the gauge field strengths above and in \cite{CCLPI}   differs from the standard one  by a factor of $\sqrt{2}$. }:
\bea
{\cal L}_4 &=& R\, {*{\bf 1}} - \frac{1}{2} {*d\varphi_i}\wedge d\varphi_i 
   - \frac{1}{2} e^{2\varphi_i}\, {*d\chi_i}\wedge d\chi_i - \frac{1}{2} e^{-\varphi_1}\,
( e^{\varphi_2-\varphi_3}\, {*  F_{1}}\wedge   F_{ 1}\nn\cr
 &+& e^{\varphi_2+\varphi_3}\, {*   F_{ 2}}\wedge   F_{ 2}
  + e^{-\varphi_2 + \varphi_3}\, {*  {\cal F}_1 }\wedge   {\cal F}_1 + 
     e^{-\varphi_2 -\varphi_3}\, {* {\cal F}_2}\wedge   {\cal F}_2)\nn\\
&-& \chi_1\, (  F_{1}\wedge  {\cal F}_1 + 
                   F_{ 2}\wedge  {\cal F}_2)\,,
\label{d4lag}
\eea
where the index $i$ labelling the dilatons $\varphi_i$ and axions $\chi_i$
ranges over $1\le i \le 3$.  The four U(1)  field strengths can be written in 
terms of potentials as
\bea
  F_{ 1} &=& d   A_{1} - \chi_2\, d {\cal A}_2\,,\nn\cr
  F_{ 2} &=& d  A_{ 2} + \chi_2\, d {\cal A}_1 - 
    \chi_3\, d   A_{ 1} +
      \chi_2\, \chi_3\, d  {\cal A}_2\,,\nn\cr
  {\cal F}_1 &=& d  {\cal A}_1 + \chi_3\, d  {\cal A}_2\,,\nn\cr
  {\cal F}_2 &=& d  {\cal A}_2\,.
\eea
 The four-dimensional theory can be obtained from six-dimensions, by 
reducing on a two-torus the following action  (See, e.g., \cite{CCLPI} and references therein.):
\ben
{\cal L}_6 = R\, {*{\bf 1} } -\frac{1}{2}{*d\phi}\wedge {d\phi}-  \frac{1}{2} e^{-\sqrt2 \phi}\, {*F_{(3)}}\wedge F_{(3)}
\label{d6lag}
\een
The above six-dimensional action is a consistent truncation of the  Neveu-Schwarz Neveu-Schwarz sector of toroidally 
compactified superstring theory. 

In the following we shall employ  the form of the gauge potentials $A_I$  $(I=1,2,3,4$) which define: $* F_1$, $F_2$,  $* {\cal F}_1$  and  $ {\cal F}_2$, respectively. (In the static case these gauge potentials  all correspond to electric fields.)
 The four-charge rotating  solution \cite{CYI}  with all the sources  explicitly displayed  was  given in   \cite{CCLPI} \footnote{Black hole solutions of the Lagrangian density (\ref{d4lag}) are generating solutions of ${\cal N}=4$ and ${\cal N}=8$  supergravity theory, which can be obtained as a  toroidal compactification on an effective heterotic string theory and Type IIA superstring theory, respectively. The full set of solutions can be obtained by acting with a subset of  respective $\{S,T\}$-  and $U$- duality transformations. (See e.g., \cite{CH}.)} 
. Here we display  the metric, only:
\bea
d{ s}^2_4 & = -{  \Delta}^{-1/2}_0 { G} ( d{ t}+{ {\cal  A}})^2 + { \Delta}^{1/2}_0 
\left( {d{ r}^2\over { X}} + d\theta^2 + {{ X}\over{  G}} \sin^2\theta d\phi^2\right)~,\label{metricg4d}
\eea
where 
\bea
{ X} & =& { r}^2 - 2{ m}{ r} + { a}^2~,\cr
{ G} & = &{ r}^2 - 2{ m}{ r} + { a}^2 \cos^2\theta ~, \cr
{ {\cal A}} & =& {2{ m} { a} \sin^2\theta \over { G}}
\left[ ({ \Pi_c} - { \Pi_s}){  r} + 2{ m}{ \Pi}_s\right] d\phi~,
\eea
and
\bea
{ \Delta}_0 =&& \prod_{I=1}^4 ({ r} + 2{ m}\sinh^2 { \delta}_I)
+ 2 { a}^2 \cos^2\theta [{ r}^2 + { m}{ r}\sum_{I=1}^4\sinh^2{ \delta_I}
+\,  4{ m}^2 ({ \Pi}_c - { \Pi}_s){ \Pi}_s  \cr && -  2{ m}^2 \sum_{I<J<K}
\sinh^2 { \delta}_I\sinh^2 { \delta}_J\sinh^2 { \delta}_K]
+ { a}^4 \cos^4\theta~.   
\eea
We are employing the following  abbreviations:
\ben
{ \Pi}_c \equiv \prod_{I=1}^4\cosh{ \delta}_I 
~,~~~ { \Pi}_s \equiv  \prod_{I=1}^4 \sinh{ \delta}_I~.
\een
The solution is parameterised by the bare mass parameter $m$, the rotational parameter $a$ and four charge parameters $\delta_I$ $(I=1,2,3,4)$.  The solution is written as a U(1) fibration over the three dimensional base, independent of the charge parameters, and  the warp factor denoted by  $\Delta_0$.

In the static case one sets $a=0$ and the solution simplifies significantly \cite{CYIII}:
\bea
d{ s}^2_4 & = -{  \Delta}^{-1/2}_{0s} { X}  d{ t}^2 + { \Delta}^{1/2}_{0s} 
\left( {d{ r}^2\over { X}} + d\theta^2 + \sin^2\theta d\phi^2\right)~,\label{metrics4d}
\eea
where
\bea
{ X} & =& { r}^2 - 2{ m}{ r} ~, \\
{ \Delta}_{0s}& = &\prod_{I=1}^4 \, ( r + 2m\sinh^2 \delta_I)~,
\eea
and  the scalar fields and the gauge potentials take the form:
\bea
&&\chi_i=0\, , \quad e^{\varphi_1} = \left[\frac{(r+2m\sinh^2\delta_1)(r+2m\sinh^2\delta_3)}{(r+2m\sinh^2\delta_2)(r+2m\sinh^2\delta_4)}\right]^{\frac{1}{2}}\,,\nn\\
&&e^{\varphi_2} = \left[\frac{(r+2m\sinh^2\delta_2)(r+2m\sinh^2\delta_3)}{(r+2m\sinh^2\delta_1)(r+2m\sinh^2\delta_4)}\right]^{\frac{1}{2}}\,,\quad
e^{\varphi_3} = \left[\frac{(r+2m\sinh^2\delta_1)(r+2m\sinh^2\delta_2)}{(r+2m\sinh^2\delta_3)(r+2m\sinh^2\delta_4)}\right]^{\frac{1}{2}}\,,\quad\nn\\
&& A_{I} = \frac{2m\sinh\delta_I\cosh\delta_I}{r+ 2m \sinh^2\delta_I}\, dt, \quad{\rm(I=1,2,3,4)} \, .~ \label{sources4d}
\eea

\subsection{Static Case}

We shall first demonstrate the scaling limit, leading to the subtracted geometry of the general {\it static} solution. 
We perform a scaling limit on the static solutions (\ref{metrics4d})-(\ref{sources4d})   where without loss of generality we take three equal charges  and the fourth one different  by defining
  $* F_1= F_2= * {\cal F} _1\equiv F\, $  and  $ {\cal F}_2\equiv {\cal F}$.  \footnote{While one can in principle perform a scaling limit with three unequal large charges
   $Q_i$ ($I=1,2,3$),  by  replacing in the scaling limit  (\ref{scalings})   $Q\to (\Pi_{I=1}^3Q_I)^{1\over 3}$,   appropriate powers of $Q_I$ in the scalar fields $\varphi_i$ ($i=1,2,3$)  and  gauge field strengths  $* F_1, F_2,  * {\cal F} _1$
  can be removed  without loss of  generality,  resulting in   the same  gauge choice for sources (\ref{ssources}).}
   We use the ``tilde'' notation for all the variables, with the choice of charge parameters ${\tilde \delta}_1={\tilde \delta_2}={\tilde \delta_3}\equiv {\tilde \delta}$ and 
  $\tilde \delta_4\equiv {\tilde \delta_0}$. 
  We take the  following scaling limit with   $\epsilon \to 0$: \bea
&&{\tilde r}= r  \epsilon,   \quad{\tilde t}= t{\epsilon^{-1}},  \quad  {\tilde m}=  m \epsilon\, , \ \cr
&&2{\tilde m} \sinh^2 {\tilde \delta} \equiv Q = {2m}{\epsilon^{-1/3}} (\Pi_c^2-\Pi_s^2)^{1/3},   \quad \sinh^2{\tilde \delta}_0=\frac {\Pi_s^2}{\Pi_c^2-\Pi_s^2}\, ,   \
\label{scalings} \eea
where  the ``tilde''  coordinates and parameters of the scaled  solution are related to those of the subtracted geometry for  the  four-charge static black hole. In the latter case the metric  of the (unsbtracted) black hole solution is of  the 
 form (\ref{metrics4d}), but with the subtracted geometry  the metric (\ref{metrics4d}) is the same, except for  the warp factor: 
\ben
\Delta_{0s}\to \Delta_s= (2m)^3 r (\Pi_c^2-\Pi_s^2) + (2m)^4 \Pi^2_s\ .
\een
The  sources  supporting this geometry are obtained by taking the scaling limit (\ref{scalings})   in  (\ref{sources4d}) (with ``tilde'' coordinates and parameters):
\bea
&&\chi_1=\chi_2=\chi_3=0 , \ \ e^{\varphi_1} =e^{\varphi_2} =e^{\varphi_3} = \frac{Q^2} {\Delta_s^{1\over 2}}, \cr
&&A=-\frac{r}{Q}\, dt , \ \ {\cal A}= \frac{Q^3 (2m) \Pi_c\Pi_s }{(\Pi_c^2-\Pi_s^2)\Delta_s}\, dt ,
\eea
resulting in field strengths:
\bea&&F_{t\, r}= \frac{1}{Q}, \quad {\cal F}_{t\, r}= \frac{Q^3(2m)^4  \Pi_c\Pi_s }{\Delta_s^2}\, .
\eea
The  (formally infinite) factors of Q can be removed  with  sources taking the form:
\bea
&&\chi_1=\chi_2=\chi_3=0 , \ \ e^{\varphi_1} =e^{\varphi_2} =e^{\varphi_3} = \frac{(2m)^2} {\Delta_s^{1\over 2}}\, ,
\cr
&&A=-\frac{r}{2m}\, dt , \ \ {\cal A}= \frac{(2m)^4 \Pi_c\Pi_s }{(\Pi_c^2-\Pi_s^2)\Delta_s}\, dt ,\label{ssources}
\eea
with electric field strengths:
\bea
&&F_{t\, r}= \frac{1}{2m}\ , \quad
{\cal F}_{t\, r}= \frac{(2m)^7 \Pi_c\Pi_s }{\Delta_s^2}\, .
\eea
%
The result for sources is  the same  (up to a  gauge choice) as the one  obtained in \cite{CL11II} by directly solving Einstein equations with  the subtracted geometry static metric.
Note that the sources supporting this geometry are those of  the minimal supergravity in five dimensions,  where $F$  is the Maxwell  field strength of the  five-dimensional theory and  ${\cal F}$ the Kaluza-Klein field strength.

\subsection{General Rotating Case}

We now proceed with obtaining a subtracted geometry for a general four-charge {\it rotating} black hole, whose original metric (\ref{metricg4d}) was displayed at the beginning of this section.

In order to  track the effects associated with the rotational parameter $a$ in the scaling limit we display explicitly the metric and the sources for 
the solution  (transcribed from  \cite{CCLPI})  with three equal charges  and the fourth one different, i.e., by  again choosing, without loss of generality,  the gauge potentials $A_1=A_2=A_3\equiv A $ for gauge field strengths    $* F_1= F_2= * {\cal F} _1\equiv F\, $  and  $A_4\equiv {\cal A}$ for  $ {\cal F}_2\equiv {\cal F}$.  The metric is written 
as  above (\ref{metricg4d}), but with  all the quantities taken with ``tilde'' notation and  denoting ${\tilde \delta}_1={\tilde \delta}_2={\tilde \delta}_3\equiv {\tilde \delta}$ and ${\tilde \delta}_4\equiv {\tilde \delta}_0$.

The  scalar fields are given by:
\bea
&&\chi_1=\chi_2=\chi_3= \frac{2{\tilde m}\, {\tilde a}\cos\theta\, \cosh{\tilde \delta} \sinh {\tilde \delta}(\cosh{\tilde \delta}\sinh{\tilde \delta}_0-\sinh {\tilde \delta}\cosh {\tilde \delta} _0)}{({\tilde r}+2{\tilde m}\sinh^2 {\tilde \delta})^2+ {\tilde a}^2\cos^2\theta},\cr
&&e^{\varphi_1} =e^{\varphi_2} =e^{\varphi_3} = \frac{({\tilde r}+2{\tilde m}\sinh^2 {\tilde \delta})^2+ {\tilde a}^2\cos^2\theta}{{\tilde {\Delta}_0}^{1\over 2}}\, ,
\eea
and the gauge potentials by:
\bea
A = && \frac{2\tilde m}{\tilde \Delta_0}\, \{
  [({\tilde r}+2{\tilde m}\sinh^2 {\tilde \delta})^2({\tilde r}+2{\tilde m}\sinh^2 {\tilde \delta}_0)+{\tilde r}{\tilde a}^2\cos^2\theta][\cosh {\tilde \delta}\sinh {\tilde \delta} \, d{\tilde t} \cr
 &&  -{\tilde a}\sin^2\theta
\cosh {\tilde \delta}\sinh {\tilde \delta}(\cosh {\tilde \delta}\cosh {\tilde \delta}_0-\sinh {\tilde \delta}\sinh {\tilde \delta}_0)\, d\phi] \cr
&&+ 2{\tilde m} \,{\tilde a}^2\cos^2\theta \,[e \, d{\tilde t} - {\tilde a}\sin^2\theta\sinh^2 {\tilde \delta}\cosh {\tilde \delta}\sinh {\tilde \delta}_0\,
  d\phi]\}\, , \cr
{\cal A} = &&\frac{2\tilde m}{ {\tilde \Delta_0}}\, \{
  [({\tilde r}+2{\tilde m}\sinh^2 {\tilde \delta})^3+{\tilde r}{\tilde a}^2
 \cos^2\theta][\cosh{\tilde \delta}_0\sinh{\tilde \delta}_0 \, d{\tilde t}\cr &&  -{\tilde a}\sin^2\theta(
\cosh^3{\tilde \delta}\sinh{\tilde \delta}_0 -\sinh^3{\tilde \delta}\cosh {\tilde \delta}_0)\, d\phi] \cr
&&+ 2{\tilde m} \,{\tilde a}^2\cos^2\theta \,[e_0 \, d{\tilde t} -{\tilde a} \sin^2\theta\sinh^3 {\tilde \delta}\cosh {\tilde \delta}_0\,
  d\phi]\}\,.
  \eea 
Here:
\bea 
e&=&\sinh^2 {\tilde \delta}\cosh^2 {\tilde \delta}\cosh {\tilde \delta}_0\sinh {\tilde \delta}_0(\cosh^2 {\tilde \delta}+\sinh^2 {\tilde \delta})\cr &&-\sinh^3 {\tilde \delta}\cosh {\tilde \delta} (\sinh^2 {\tilde \delta}+2\sinh^2 {\tilde \delta}_0+2\sinh^2 {\tilde \delta}\sinh^2 {\tilde \delta}_0),\cr
e_0&=&\sinh^3 {\tilde \delta}\cosh^3 {\tilde \delta}(\cosh^2 {\tilde \delta}_0+\sinh^2 {\tilde \delta}_0)-\sinh {\tilde \delta}_0\cosh {\tilde \delta}_0(3\sinh^4 {\tilde \delta}+2\sinh^6 {\tilde \delta})\, .
\eea
%
%
Again,  we  take the  scaling limit   (\ref{scalings}),  and furthermore we   take for the rotational parameter:
\ben
 {\tilde a}= a   \epsilon\, .
 \label{scalinga} 
 \een
In terms of new coordinates and parameters  the metric  takes the form   (\ref{metricg4d}), where only the warp factor changes:
\ben 
\Delta_0\to \Delta 
= (2m)^3 r (\Pi^2_c - \Pi^2_s) + (2m)^4 \Pi^2_s - (2m)^2 a^2(\Pi_c-\Pi_s)^2 \cos^2\theta~.
\label{deltag}\een
This geometry with the subtracted warp factor is sourced by the scalars:
\ben
\chi_1=\chi_2=\chi_3=-\frac{2ma(\Pi_c-\Pi_s)cos\theta}{Q^2} , \ \ e^{\varphi_1} =e^{\varphi_2} =e^{\varphi_3} = \frac{Q^2} {\Delta^{1\over 2}}\, ,
\een
and  the gauge potentials:
\bea
A=&&-\frac{r}{Q}dt+\frac{(2m)^2 a^2[2m\Pi_s^2 -r (\Pi_c-\Pi_s)^2]\cos^ 2\theta}{Q\Delta}dt \cr
&&-\frac{2m\, a(\Pi_c-\Pi_s)\sin^2\theta }{Q}\left(1+\frac{(2m)^2a^2(\Pi_c-\Pi_s)^2\cos^2\theta}{\Delta}\right)d\phi\, ,
\cr
{\cal A}=&& \frac{ Q^3[(2m)^2 \Pi_c\Pi_s + a^2 (\Pi_c-\Pi_s)^2\cos^2\theta]}{2m(\Pi_c^2-\Pi_s^2)\Delta}\, dt \, + \frac{Q^32m\,a(\Pi_c-\Pi_s)\sin^2\theta}{\Delta}   \, d\phi\,,
\eea
resulting in  field strengths
with  both electric and magnetic components.
%
The  (formally infinite) factors of Q can  again be removed from gauge potentials  by removing   corresponding factors from scalar fields, and thus the sources take the  canonical form:
\bea
&&\chi_1=\chi_2=\chi_3=-\frac{a(\Pi_c-\Pi_s)\cos\theta}{2m} , \ \ e^{\varphi_1} =e^{\varphi_2} =e^{\varphi_3} = \frac{(2m)^2} {\Delta^{1\over 2}}\, ,\label{canonicalsc}
\eea
\bea
A=&&-\frac{r}{2m}dt+\frac{(2m) a^2[2m\Pi_s^2 -r (\Pi_c-\Pi_s)^2]\cos^ 2\theta}{\Delta}dt\, \cr\
&&-{ a(\Pi_c-\Pi_s)\sin^2\theta}(1+\frac{(2m)^2a^2(\Pi_c-\Pi_s)^2\cos^2\theta}{\Delta})\, d\phi ,
\cr
{\cal A}= &&\frac{ (2m)^4 \Pi_c\Pi_s + (2m)^2a^2 (\Pi_c-\Pi_s)^2\cos^2\theta}{(\Pi_c^2-\Pi_s^2)\Delta}\, dt \, + \frac{(2m)^4a(\Pi_c-\Pi_s)\sin^2\theta}{\Delta}   \, d\phi\, ,\label{gaugeps}
\eea
 Again, the sources supporting this geometry are those of 
 the minimal supergravity in five-dimensions, where $F$ and  ${\cal F}$  are the five-dimensional Maxwell and  Kaluza-Klein field strengths, respectively.

The scaling limits (\ref{scalings}),(\ref{scalinga})  are reminiscent of the near-BPS dilute gas approximation \cite{MS},   which were generalized to rotating four-dimensional black holes in \cite{CL99I}.  
  As a  natural consequence,  the  subtracted geometry of general black holes  is  a Kaluza-Klein coset of  $AdS_3 \times 4 S^2$ just as
 in the dilute-gas approximation \cite{CL99I}. Furthermore, there  is an analogous 
microscopic interpretation  in terms of two-dimensional conformal field theory of a long rotating string, which was addressed in \cite{CL11II}.  

In Appendix  A we show that the subtracted geometry of general five-dimensional black holes can be obtained as an analogous
scaling limit, reminiscent of the  near-BPS dilute gas approximate for five-dimensional rotating black holes \cite{CL99II}, resulting in the Kaluza-Klein coset of $AdS_3\times S^3$, and analogous microscopic interpretation via a conformal  field theory of a long  rotating string, studied in \cite{CL11I}. In Section 4 we further  analyse  geometric properties of  emerging  Kaluza-Klein cosets.
\subsection{Subtracted Geometry as a Harrison 
Transformation }

In this Subsection we demonstrate that the subtracted geometry can be obtained as a  ic Harrison transformation on the original black hole solution. For the sake of simplicity and in order to demonstrate the procedure we shall present the details  for 
the Schwarzschild black hole, only. In this case it is sufficient to employ  the Einstein-Dilaton-Maxwell Lagrangian density, with the dilation coupling $\alpha={1\over \sqrt{3}}$, which is a  consistent truncation of the Lagrangian density (\ref{d4lag}) with $\chi_i=0$, 
 $\varphi_i=\varphi_2=\varphi_3\equiv
-\frac{2}{\sqrt{3}}\phi$, $*F_1=F_2=*{\cal F}_1\equiv \sqrt{\frac{2}{3}} F$  and ${\cal F}_2=0$.  [Of course for the multi-charged rotating black holes  one has to employ the full   {\cal N}=2 supergravity Lagrangian  density (\ref{d4lag}).]

We begin by considering static solutions to 
general Einstein-Dilaton-Maxwell equations with the general dilation coupling $\alpha$. The Lagrangian density  is
\footnote{We choose the  units in which $4 \pi G=1$. Note that  in the   Lagrangian density  (\ref{d4lag})   $16\pi G= 1$ and the field strengths differ  by a factor of $\sqrt{2}$. }.:
\ben
\sqrt{-g } \Bigl ( \frac{1}{4} R  - \half (\p \phi) ^2 
- \frac{1}{4} e^{-2\alpha \phi} F^2 \bigr ) \,. \label{d4edm}
\een
Making the Ansatz
\ben
ds ^2 =-e^{2U} dt^2 + e^{-2U} \gamma_{ij}dx^i dx^j \,,\qquad F_{i0}=\p_i \psi 
\een
we obtain an effective action density  in three  dimensions of the form
\ben
\sqrt{\gamma} \Bigl( R(\gamma _{ij} ) -2 \gamma ^{ij} \Bigl( \p_i U\p_jU    + \p_i \phi \p_j  \phi
- e^{-2U} e^{-2\alpha \phi}  \p_i \psi \p_j  \psi      \Bigr) \Bigr )  
\een
Defining
\ben
x\equiv  \frac{U+\alpha  \phi}{\sqrt{1+\alpha^2}} \,,\qquad y\equiv \frac{-\alpha U+ \phi}{\sqrt{1+\alpha^2}}
\,, \een
the effective action density becomes 
\ben
\sqrt{\gamma} \Bigl 
(R(\gamma _{ij} ) -2 \gamma ^{ij} \Bigl( \p_i x \p_j x   + \p_i y \p_j  y
- e^{- {2}{\sqrt{1+\alpha ^2}}x }   \p_i \psi \p_j  \psi      \Bigr) \Bigr ) \,.  \label{d3lag}
\een
Evidently we can consistently set $y=0$ and we obtain 
a sigma model, whose fields $x,\psi$ map into
the  target $SL(2,\R)/SO(1,1)$,  coupled to three dimensional
Einstein gravity. The non-trivial action of an $SO(1,1)$ subgroup of
$SL(2,\R)$ is called a  Harrison  transformation.

More concretely, and following \cite{Yaz} but making some  changes 
necessitated by considering a reduction on
time-like, rather than a space-like Killing vector we 
define a matrix (See also, e.g., \cite{Gal} and references therein.):
\bea
P &= &
e^{-\sqrt{1+\alpha^2}(x+y)} \pmatrix {e^{2\sqrt{1+\alpha^2}x} - (1+\alpha^2) \psi ^2 & - \sqrt{1+\alpha ^2} \psi \cr - \sqrt{1+\alpha ^2} \psi &-1\cr}\, , 
\eea
so that
\ben
P= P^T\,,\quad
  \det  P
  = -e^{-2\sqrt{1+\alpha^2}y} \,. \label{properties}
\een
Taking  $H\in SO(1,1)$
which acts   on $P$ as 
\ben
P \rightarrow H P H^T\, , 
\een 
it preserves  not only the properties (\ref{properties})  
but also  the Lagrangian density (\ref{d3lag}) which can be cast in the form:
\ben
\sqrt{\gamma} \Bigl( R(\gamma_{ij})  + \frac{1}{1+\alpha^2}  \gamma^{ij}  {\rm Tr}
(\p_i
P \p_j 
P^{-1} )  \Bigr) .
\een
It is straightforward to show  that a Harrison transformation:
 \ben H=\pmatrix{ 1& 0 \cr \beta & 1} \,  ,
 \label{harr} 
 \een
corresponds to:
\bea y'&=&y\,  , \quad \quad    e^{\sqrt{1+\alpha^2} x'}= \Lambda^{-1} e^{\sqrt{1+\alpha^2} x}\, , \cr
  \psi'&=&\Lambda^{-1}[\psi +\frac{\beta}{\sqrt{1+\alpha^2}}(e^{2\sqrt{1+\alpha^2}x} -(1+\alpha^2)\psi^2)]\, ;  \   \Lambda=(\beta\psi+1)^2 -\beta^2e^{2\sqrt{1+\alpha^2}x} \, .\label{harrp}
\eea
Note, this transformation can also  be determined as an analytic continuation  of  transformations given in Section 2 of \cite{Yaz}. 
A Harrison transformation in the limit of an infinite boost 
corresponds to $\beta\to 1$. One may verify that  (\ref{harr})  with $\beta\to 1$ 
in the Einstein-Maxwell gravity ($\alpha =0$) 
takes the Schwarzschild metric  to the Robinson-Bertotti one. This type of transformation was employed recently in \cite{Bertini}. For another work, relating   the Schwarzschild geometry
 to $AdS_2\times S^2$, see  \cite{Franzin}.
%

In   the case of $\alpha={1\over\sqrt{3}}$,   we shall act with (\ref{harr})  on the Schwarzschild solution with
$e^{2U}=1-{{2m}\over r}$, $\phi=0$, $\psi=0$. The transformation (\ref{harrp}) with $\beta=1$  results in $\Lambda=\frac{2m}{r}$, and  the  metric  (\ref{metrics4d})  with the  subtracted geometry 
warp factor:\ben\Delta_{s0}=r^4\to \Delta_s=(2m)^3 r\, ,\een
and  the scalar field and the electric field strength :
\ben
e^{-{{2\phi}\over {\sqrt{3}}}}=\sqrt{{2m\over r}}\, , \, \quad \sqrt{\frac{2}{3}}F_{t\,r}={1\over{2m}}\, ,
\een
i.e., this is  the static  subtracted geometry  of Subsection 2.1,  with $\Pi_c=1,\ \Pi_s=0$.

The subtracted geometry for the Kerr spacetime  can  be obtained by reducing the spacetime on the  time-like Killing vector and acting  on the Kerr black hole with an infinite boost Harrison transformation  for  Lagrangian density (\ref{d4lag}), where we set $\chi_1=\chi_2=\chi_3\equiv \chi $, $\varphi_1=\varphi_2=\varphi_3\equiv \frac{2}{\sqrt 3}\phi$, $*F_1=F_2=*{\cal F}_1\equiv \sqrt{\frac{2}{3}} F$ and ${\cal F}_2=\sqrt{2}{\cal F}$, i.e.  an Einstein-Dilaton-Axion gravity with two $U(1)$ gauge fields and respective dilaton couplings  $\alpha_1 ={1\over \sqrt{3}}$ and $\alpha_2=\sqrt{3}$.
 The subtracted geometry of the multi-charged rotating black holes  is expected to arise  as a ic Harrison transformation on a rotating  charged black solution of (\ref{d4lag}).  This has recently been confirmed \cite{Vir}.

These results  demonstrate that the subtracted geometry is  a solution of the same theory as the original black hole. Furthermore the original black hole and  the subtracted geometry clearly lie in the same duality orbit, specified above and  passing through the original black hole. Thus any physical property of the original black hole solution which is invariant under the duality transformation of M-theory  remains the property  of the subtracted geometry. For example the area of the horizon is unchanged.
 %

\section{Asymptotically Conical Metrics}
 
The scaling  limit, or equivalently  the  subtraction process, alters the 
environment that our black holes find themselves in \cite{CL11I,CL11II}.  In fact
the subtracted geometry metric is  asymptotically of the  form  
\ben
ds ^2 = -\bigl( \frac{R}{R_0}\bigr )  ^{2p} dt ^2 + B^2 dR ^2 + R^2 \bigl( d \theta ^2 + \sin ^2 \theta ^2 d \phi ^2  )
\bigr ) \label{cone} 
\een
with $R_0$ a constant, $B=4$ and $ p=3$.
In general, metrics with asymptotic form (\ref{cone})  may be referred as 
{\it Asymptotically Conical} (AC). The spatial metric is conical because the
radial distance $BR$ is a non-trivial multiple
of the area distance $R$. Restricted to the equatorial plane
the spatial metric is that of a  flat two-dimensional cone
\ben
ds ^2 _{\rm equ} = B^2 dR^2 +R^2 d \phi ^2 
\een  
with deficit angle 
\ben
2 \pi (1- \frac{1}{B}) = 8 \pi \eta ^2 \,. 
\een
A characteristic feature of conical metrics of the form (\ref{cone}) is that
they admit a {\it Lifshitz scaling}.  That is they admit a {\it homothety},
a diffeomorphism under which the pulled back metric goes into
a constant  multiple of itself. In our case the homethety is
\ben
R \rightarrow \lambda R \,,\qquad t \rightarrow  \lambda ^{1-p} t \,, \qquad
\Rightarrow \qquad ds ^2 \rightarrow \lambda^2 ds ^2 
\een 
or if one introduces isotropic Cartesian coordinates ${\bf x} = R^B (\sin \theta \cos \phi,\sin \theta \sin \phi,\cos \theta)$ which render
the spatial metric conformally flat, we have scaling under,
\ben
\bx \rightarrow \lambda \bx   \,, \qquad t \rightarrow  \lambda ^z t\,, 
\een
where the difference of the Lifshitz scaling exponent $z$ from unity
is a measure of how space and time scale differently. 
We have 
\ben
z= \frac{1-p}{B} 
\een
and so $z=-\half$ in our case.  The nonstandard scaling of time 
is reminiscent of the Lifshitz symmetry that has recently been developed for
applications of holography to condensed matter systems. (See, e.g., \cite{KachruYH,DanielssonGI,BertoldiVN} and references therein.).

 Since asymptotically Conical (AC)   metrics of the type (\ref{cone})  may be unfamiliar to the string theory community, we briefly  recall  some of their properties and give some examples.  
AC metrics typically arise when the energy density $T_{00}$  
of a static  four dimensional 
spacetime falls off as $T_{00} \rightarrow \frac{\eta^2}{R^2}$ \footnote{These 
components in a pseudo-orthonormal frame.  
The coordinate $R$ is the  area
distance. We  use throughout signature $-+++$ and 
 units in which Newton's constant $G=1$.}. 
Because   of this slow fall off
the metric  cannot have finite total energy and  
cannot be {\it asymptotically flat} (AF).   At  large distances such  
spacetime metrics typically  take the  form of (\ref{cone})  near infinity.

Examples where (\ref{cone}) is exact    are  
\begin{itemize}
\item $p= \frac{2\gamma}{1+ \gamma}$ , $ B= 
\frac{\sqrt{1+ 6 \gamma + \gamma ^2}}{1+\gamma}$, gives   
{\it Bisnovatyi-Kogan Zeldovich's  gas sphere} \cite{B-KZ,B-KT} 
Here, $\gamma$ is the constant ratio of pressure to density  of. 
the gas for which
\ben
P \propto  \frac{\gamma ^2}{1+6 \gamma + \gamma ^2} \frac{1}{2 \pi R^2} 
 \een 
\item $p=0$ and $B= \sqrt{1-8\pi  \eta ^2 } $, gives the 
{Barriola-Vilenkin  Global Monopole} \cite{Barriola}.
The source if this metric is an $SO(3)$ non-linear
sigma model with the Higgs field of constant magnitude $\eta$
in the hedgehog configuration. 

\item The near horizon geometry of an extreme black hole
in Einstein-Dilaton-Maxwell gravity  with coupling constant $\alpha$   
has  $p = \frac{1}{\alpha^2} $, $z=1- \frac{1}{\alpha^2} $ 
and $B^2 = \frac{1+ \alpha^2}{\alpha^2} $. Our case corresponds 
to $\alpha^2 =\frac{1}{3}$. 
\end{itemize}

Examples of  metrics which are asymptotically conical 
rather than being exactly conical are that of   
a black hole containing a global monopole \cite{Barriola}
possibly with a  magnetic (or   electric) 
charge \cite{Gibbons} for which
\ben
ds^2 = - \bigl (1 - 8\pi  \eta ^2 -\frac{2m}{R} + \frac{P^2}{R^2}  
\bigr )dt^2  + \frac{dR^2}{1 - 8\pi  \eta ^2 -\frac{2m}{R} + \frac{P^2}{R^2}  }  +  R^2 \bigl( d \theta ^2 + \sin ^2 \theta  d \phi ^2  )\,.
\een
One may also combine a global monopole with a Kaluza-Klein monopole
\cite{Banerjee} with four-dimensional metric 
\ben
ds^2 = - \bigl (1 - 8\pi  \eta ^2 -\frac{2m}{R}    
\bigr )^\half dt^2  + 
\frac{dR^2}{\bigl ( 1 - 8\pi  \eta ^2 -\frac{2m}{R} \bigr)
^\half  }  +             \bigl (1 - 8\pi  \eta ^2 -\frac{2m}{R}    
\bigr ) ^\half           R^2 \bigl( d \theta ^2 + \sin ^2 \theta  d \phi ^2 \bigr ) \,.
\een
A rather difference example is obtained by dimensional reduction
of  an $8$-dimensional ultra static metric whose spatial
metric is an asymptotically a cone over $S^3 \times S^3$  with holonomy
$G_2$ \cite{Hartnoll}. This has $B^2= \frac{4}{3}$ and $p=\frac{2}{3}$.

\subsection{Flat Spacetime Analogs}

In addition to having energy-densities falling off as $R^{-2}$,  the subtracted geometries have
the property that scalar fields do not tend to a constant at spatial infinity (\ref{canonicalsc}), but they run logarithmically at large $R$.  Despite of  this  fact the total charges are finite. 
In fact this somewhat unfamiliar behavior  can occur in general Einstein-Dilaton-Maxwell theory. Moreover, it is possible to take a limit in which gravity decouples and one discovers
very similar behaviour in  Dilaton-Maxwell theory  \cite{Wells}.  
Although the relevant solution has infinite energy in flat spacetime,  
the Maxwell and dilaton fields are perfectly regular
outside the origin and they possess finite total
electric or magnetic charge. Since their properties
resemble many of their  fully self-gravitating cousins
we provide below a brief self-contained derivation.  
Of  particular interest is the  fact that   the dilaton $\phi$
in these theories  provides
a  spacetime dependent abelian gauge coupling constant $g$ \footnote{
Equivalently one has  spacetime dependent magnetic permeability
$\mu =g^2 $ and electric permittivity $\epsilon = g^{-2} $.}:  
\ben
g= e^{\alpha \phi}\, ,  
\een
which in the static solutions 
 ``runs'' from zero to infinity (magnetic case) or infinity to zero
(electric case) as the radius $r$ runs from zero to infinity.
A similar running is seen in the fully self-gravitating 
solutions. The details are as follows. 

The flat spacetime Lagrangian density:
\ben
4 \pi {\cal L} = -\half (\p \phi)^2 - \frac{1}{4} e^{-2 \alpha \phi} F_{\mu \nu} ^2  \, , 
\een
leads to the following equation for the static dilaton field:
\ben
\babla ^2 \phi -\alpha e^{-2\alpha \phi} \bE ^2 + \alpha e^{2\alpha \phi} \bB ^2 \,
=0 \, , \een 
and the electric  and magnetic field Ans\"atze:
\ben
D_r = E_r e^{-2\alpha \phi} = \frac{ Q}{r^2} \,, \qquad
B_r = \frac{ P}{r^2}\, .
\een
Thus
\ben
\babla ^2 \phi = \frac{1}{r^4} \alpha \bigl( Q^2 e^{2 \alpha \phi} - P^2 e^{-2a \phi}  \bigr )  \, .
\een
Defining  $t\equiv \ln r$, we find:
\ben
\ddot \phi =   \alpha\bigl( Q^2 e^{2 \alpha \phi} - P^2 e^{-2\alpha \phi} \bigr ) \, .
\een
If $P=0$ we try
\ben
\phi = A \ln r +B \, ,
\een
and find
\ben
A=\frac{1}{\alpha} =\alpha Q^2 e^{2 \alpha B}    \, .
\een 
If $Q=0$ we try
\ben
\phi = A \ln r +B \, .
\een
and find
\ben
A=- \frac{1}{\alpha} = -\alpha  P^2  e^{-2 \alpha B}   \, . 
\een
In both cases
\ben
4 \pi T_{00} = \frac{1}{\alpha^2 r^2} \, .
\een

\subsection{Confining Properties}

It is helpful when discussing  black hole thermodynamics to consider black hole confined in a box. 
One widely used example of such a box  is provided by asymptotically anti-de Sitter space-time. 
In this case the spatial geometry is asymptotically hyperbolic and the blue-shift factor $\sqrt{-g_{00}}$  increases exponentially with proper radial distance. 
This leads to confinement of massive particles and of  thermal radiation. By  Tolman's red-shifting formula for thermal radiation, the temperature  
$T\propto 1/\sqrt{-g_{00}}$ and thus falls off  exponentially with proper distance. As a consequence the total energy and entropy outside any given radius  are finite.

In the following we show that subtracted geometry metrics have similar properties, thus justifying our claim that these metrics represent black holes in confining boxes. Actually, this confining property is a feature of AC metrics in general (\ref{cone}) with $p>1$.

Since  $\sqrt{-g_{00}}\propto R^p$,   the situation is qualitatively similar to the anti-de Sitter case.
More quantitatively, it is instructive to consider the motion of light-rays in this background.
 Their spatial projections are geodesics of the optical  metric 
\ben
ds^2_o= B^2 \bigl( \frac{R_0}{R} \bigr )^{2p} dR ^2 + 
R^2  \bigl( \frac{R_0}{R} \bigr )^{2p} \bigl ( d \theta ^2 + \sin ^2 \theta d \phi ^2 \bigr) 
\een
If $ \tilde R = R^{1-p} R_0 ^p$
\ben
ds ^2_o =   ( B^\prime)^2  d {\tilde R} ^2  + {\tilde R}^2   
\bigl ( d \theta ^2 + \sin ^2 \theta d \phi ^2 \bigr)  
\een
with $B^\prime = \frac{B}{|1-p|} $. The optical manifold   is  also a  
{ \it  cone over a 2-sphere}.  However, whereas if $p<1$, 
$\tilde R $  increases as $R $ increases and so spatial
infinity is at an infinite optical distance,  
if  $p>1$, then $\tilde R$ decreases as $R$-increases
and infinity is a cone $\tilde R =0$  at a finite  optical distance.  
Restricted to the equatorial plane
the optical  metric is that of a  flat two-dimensional cone
\ben
2 \pi (1- \frac{1}{B^{\prime}}) = 2 \pi (1- \frac{|1-p|}{B})   \,. 
\een
In our case, $p=3$ and so outwardly directed  light rays with $R$ 
 will, unless strictly radial,  spiral  around the 
optical cone whose vertex is at infinity
for a finite time and  return inwardly  directed. 
The strictly radial light rays will reach infinity in finite time. 
In this sense the environment acts rather like a box surrounding the
horizon in a fashion reminiscent of black holes in asymptotically AdS
spacetimes.

As far as the Tolman red shifting is concerned the temperature of thermal radiation falls of as $R^{-p}$, thus the total entropy
outside $R$ will be finite if $p>1$ which is clearly satisfied in our case with $p=3$.

\section{Symmetries of the Subtracted Geometry}

\subsection{Breaking of $(SL(2,\R) \times SL(2,\R) )/\Z_2 \times SO(3)$
by Kaluza-Klein Reduction}

All of the $4$-dimensional subtracted geometries 
considered in this paper may be obtained \cite{CL11II}
by reduction from the five-dimensional metric:
\ben ds ^2_5= ds^2_{ AdS_3} + 4 ds^2 _{S^2} \label{ads3}
\een
which is,
up to a factor, the sum of the maximally symmetric metric 
on $AdS_3 \equiv SL(2,\R)$ of unit radius  and  
the    round
metric on a unit $2$-sphere. The isometry group of $ds^2_5$ is
thus $ SO(3) \times ((SL(2, \R) \times SL(2,\R) )  /\Z_2)  $.
In the general case  the space-like Killing vector ${\partial}\over{\partial \alpha}$  whose orbits
the Kaluza-Klein reduction  is effected generates 
a one parameter subgroup  $H$ 
of  $G=  SO(3) \times ((SL(2 \R) \times SL(2,\R))/\Z_2) $
with projections in all three factors. Thus in general the quotient
$G/H$   
admits  an effective action of the centraliser of $H$ in $G$ 
which in general consists of 
just two Killing vectors $\p_t$ and $\p_\phi$.   
Nevertheless when solving for the massless wave equation
in these geometries  one discovers that the solutions may be expressed
in terms of hypergeometric functions (see, e.g., \cite{MS,CL97I}) and 
indeed the wave operator may be
expressed as a sum of the Casimir for $SL(2,\R) $ and $SO(3)$. 
Moreover, \cite{CL11II}  these Casmirs may be seen to commute 
with all of the generators of 
$ SO(3) \times ((SL(2 \R) \times SL(2,\R))/\Z_2) $. 
Since in general only two of these generators
correspond to Killing fields, this is on the face of it
puzzling. It suggest that perhaps the solutions of the massless wave
equation on the subtracted geometries  carry a representation of
 $ SO(3) \times ((SL(2 \R) \times SL(2,\R))/\Z_2) $. However
this is manifestly not true. The reason being that while solutions
of the massless wave equation on   (\ref{ads3})
do indeed carry a representation of    
$ SO(3) \times ((SL(2 \R) \times SL(2,\R))/\Z_2) $, {\sl only those solutions
invariant under the action of $H$ descend to the quotient   
$(AdS_3 \times S^2)/H$.}  Thus there is no action
of the full $ G= SO(3) \times ((SL(2 \R) \times SL(2,\R))/\Z_2) $ on
solutions of the wave equation on  the four-dimensional
spacetime  $(AdS_3 \times S^2)/H$, but just  of the centraliser
of $H$ in $G$,  which is generated by precisely 
the four-dimensional Killing fields
$\p_t$ and $\p_\phi$.    In the following Subsection we verify these results  by employing explicit expressions  for the  $SL(2,\R)\times SL(2,\R) $  and $SO(3)$ generators in terms of four-dimensional coordinates  $(t,r,\phi)$ and the fifth coordinate  $\alpha$.

\subsection{Action of  $SL(2,\R)$  Generators}
The lift of the subtracted geometry for general four-dimensional black holes to five dimensions on coordinate $\alpha$ is locally described \cite{CL11II} i as a metric (\ref{ads3})  which is a sum (up to rescaling) of the maximally symmetric metric 
on $AdS_3 \equiv SL(2,\R)$ of   and  
the    round
metric on a unit $2$-sphere. The isometry group of $ds^2_5$ is
thus $ SO(3) \times ((SL(2, \R) \times SL(2,\R) )  /\Z_2)  $. 

The $AdS$ metric can be written   in global coordinates as:
\ben
ds^2_3 = \ell_{AdS}^2 (d\rho^2 - \cosh^2\rho d\tau^2 + \sinh^2\rho d\sigma^2)~.\nn
\een

The $SL(2,\R)\times SL(2,\R)$ generators  ${\cal R}_i$ ($i=1,2,3$) and ${\cal L}_i$ ($i=1,2,3$) acting on  $AdS_3$  are  linear differential operators  which commute with each other and satisfy $SL(2,\R)$ algebras\footnote{ The operators  $\cR _i$  and $\cL _i$ 
are, in the usual way  multiples by $-{i} $ of vector fields, so as to
have real eigenvalues.} :
\ben
[{\cal R}_i, {\cal R}_j] = 2i \epsilon_{ijk} (-)^{\delta_{k3}}{\cal R}_k~~;~[{\cal L}_i, {\cal L}_j] = 2i \epsilon_{ijk} (-)^{\delta_{k3}}{\cal L}_k~,\nn
\een
and take the form:
\bea
{\cal R}_\pm &= &{\cal R}_1\pm i {\cal R}_2 = e^{\pm i(\tau+\sigma)} \left(\mp i \partial_\rho +  \coth\rho\partial_\tau + \tanh\rho\partial_\sigma\right)~,\nonumber\\
{\cal R}_3 & = &\partial_\tau + \partial_\sigma~,\label{rpm}
\eea
and ${\cal L}_i$ determined by taking $\tau\to -\tau$ . 

The $S^2$ metric is written as:
\ben
ds^2_2 = \ell_S^2 (d\Phi^2 + \sin^2\Theta d\Phi^2)~,\nn \label{s2m}
\een
where $SO(3)$ generators  ${\cal J}_i$  (j=1,2,3) on  $S^2$   satisfy the  algebra:
\ben
[{\cal J}_i, {\cal J}_j] =  \epsilon_{ijk} {\cal J}_k\, , \nn\een
and take the form:
\bea
{\cal J}_\pm &= &{\cal J}_1\pm i {\cal J}_2 = e^{\pm i\Phi} \left( \partial_\Theta +  \cot\Theta\partial_\Phi\right) ~,\\ \nonumber
{\cal J}_3 & = &\partial_\Phi ~. \nn \label{jpm}
\eea

In terms of the original coordinates
 $t,r, \theta, \phi$ and ${\tilde \alpha}$ the metric (\ref{ads3}) takes the form \cite{CL11II}:
 \bea
 ds^2_5& = &\Delta ( d{\tilde\alpha} + { \cal B})^2 + \Delta^{-1/2} ds^2_4 \\ 
 &=& - {X\over\rho}dt^2 +{dr^2\over X} +  \rho
(d{\tilde \alpha} + { {\cal A}_{\rm red}\over 2m (\Pi_c - \Pi_s)\rho}dt)^2 + d\theta^2 + \sin^2\theta d{\tilde \beta}^2~,\nn\label{lifted}
\eea
where $ds^2_4$ is the subtracted geometry four-dimensional metric (\ref{metricg4d}) with $\Delta_0\to \Delta$,  $\Delta$   defined in (\ref{deltag}), the Kaluza-Klein  gauge potential ${\cal B}$  as defined in \cite{CL11II} and it is related to ${\cal A}$ in (\ref{gaugeps})  as:
\ben {\cal B}={{\cal A}\over (2m)^3}- {{dt}\over{(2m)^3(\Pi_c^2-\Pi_s^2)}}\,,
\een
and
\bea
X&=&r^2-2mr+a^2\, ,  \nonumber\\
{\cal A}_{\rm red}&=& 2m [ (\Pi_c - \Pi_s)r + 2m \Pi_s]~, \nonumber\\
 \rho &= &{\cal A}_{\rm red}^2 - 4m^2 (\Pi_c - \Pi_s)^2 X
= 8m^3 [ r  (\Pi_c^2 - \Pi_s^2)+ 2m \Pi^2_s - {a^2\over 2m}(\Pi_c - \Pi_s)^2  ]\, , \nonumber\\
\eea
and
\ben{\tilde \beta}= \phi + 2ma(\Pi_c - \Pi_s) {\tilde\alpha}~.\nonumber
\een
The first part of the metric is $AdS_3$ with global coordinates expressed as:
\bea
\sinh^2\rho &= &{r - r_+\over r_+ - r_-}~, \ \ r_{\pm}=m\pm \sqrt{m^2-a^2}\, , \nn\\ \label{coorads0}
\nonumber
  \sigma + \tau &=&  2i\, m\sqrt{m^2-a^2}(\Pi_c-\Pi_s)\, {\tilde \alpha}\, , \nonumber\\
 \sigma - \tau &= &-i\frac{\, t}{2m(\Pi_c-\Pi_s)}-2i\, m^2(\Pi_c+\Pi_s){\tilde \alpha}\, ,\eea
and $\ell_{AdS}=2$, the  results that can  also be  inferred from \cite{CL11II}. The generators ${\cal R}_3$ and ${\cal L}_3$ of $SL(2,\R)\times SL(2,\R)$  now become:
\bea
 {\cal R}_3 & =& -i\frac{1}{m\sqrt{m^2-a^2}(\Pi_c-\Pi_s)} \partial_{\tilde \alpha} +4i\frac{m^2(\Pi_c+\Pi_s)}{\sqrt{m^2-a^2}}\partial_t\, ,\nonumber\\
 {\cal L}_3 & = &4im(\Pi_c-\Pi_s)\partial_t\, .\nonumber
\eea

We identify the  compactification coordinate $\alpha$  with $\tilde \alpha$ and expressing
the azymuthal four-dimensional angle $\phi$ in terms of  ${\tilde \alpha}$  and ${\tilde \beta}$ as:
\bea  \alpha&=&{\tilde \alpha}\,,\\ \nonumber
\phi&=&{\tilde \beta}-2ma(\Pi_c-\Pi_s){\tilde \alpha}\, ,\label{alph}
\eea
resulting in the following identification of the  differential operators:
\bea
 \partial_{\tilde \alpha}&=&\partial_\alpha-2ma(\Pi_c-\Pi_s)\partial_\phi\, , \nonumber\\
 \partial_{\tilde \beta}&=&\partial_\phi\, . \label{partiala}
 \eea
Consequently, the Cartan generators take the form:
\bea
2\pi{\cal R}_3&=& -i\frac{2\pi}{m\sqrt{m^2-a^2}(\Pi_c-\Pi_s)} \partial_{\alpha} +2i\beta_H\Omega\partial_\phi+i\beta_R\partial_t\, , \nonumber\\
2\pi {\cal L}_3 & = &i\beta_L \partial_t\, ,\label{geno}
\eea
where $\beta_H$ is the inverse Hawking temperature  and $\Omega$ the angular velocity of the original black hole  with the relation:
\ben
\beta_H\Omega=2\pi \frac{a}{\sqrt{m^2-a^2}}\, . \label{bo}
\een
The inverse Hawking temperature $\beta_H$, and that at the inner horizon $\beta_-$ can be expressed in terms of  $\beta_{L,R}$:
\ben
\beta_H=\frac{1}{2}(\beta_R+\beta_L)\, , \ \ \ \beta_-=\frac{1}{2}(\beta_R-\beta_L)\, , \label{bpm}
\een
where
\ben \beta_R=8\pi \frac{m^2}{\sqrt{m^2-a^2}}(\Pi_c+\Pi_s)\, , \ \ \ \beta_L=8\pi m(\Pi_c-\Pi_s)\, .\label{blr}
\een
On the other hand,  the $S^2$ coordinates (\ref{s2m}) are identified as:
\bea
\Theta&=&\theta\,  \nonumber\\
\Phi&=& {\tilde \beta}=\phi+2ma(\Pi_c-\Pi_s)\alpha\, , \label{TP}\eea
and $\ell_{S}=1$.  ${\cal J}_3$ generator of $SO(3)$ is just 
\ben
{\cal J}_3=\partial_\phi\, , \nonumber
\een
while  ${\cal J}_{\pm}$ have an explicit dependence on $\phi$ and $\alpha$. 

The Laplacian  on (\ref{lifted}) is a sum of $AdS_3$  and $S^2$  Laplacians which can be cast in the form  of the  quadratic Casimirs of  $SL(2,\R)$ and $SO(3)$ generators, respectively. ally:
\bea 
\ell^2\nabla_{AdS_3}^2 &= & {\cal R}_1^2 + {\cal R}_2^2 - {\cal R}_3^2~ \nonumber\\
&= &{\cal L}_1^2 + {\cal L}_2^2 - {\cal L}_3^2 \nonumber\\
&=&
{1\over \sinh 2\rho}\partial_\rho\sinh 2\rho
\partial_\rho - {1\over\cosh^2\rho}\partial^2_\tau
+ {1\over\sinh^2\rho}\partial^2_\sigma~, \\
\ell^2\nabla_{S^2}^2 &= & {\cal J}_1^2 + {\cal J}_2^2 +{\cal J}_3^2~\nonumber\\
&=&
\frac{1}{\sin\Theta}\partial_\Theta(\sin\Theta\partial_\Theta)~+
{1\over \sin^ 2\Theta}\partial^2_\Phi\, ,  \eea
The $AdS_3$ Laplacian  can be expressed in terms of  partial derivatives $\partial_r$, $\partial_t$ and $\partial_{\tilde \alpha}=\partial_\alpha-2ma(\Pi_c-\Pi_s)\partial_\phi$, while the  coefficients in front of derivatives depend on $r$, only. Similarly, the Laplacian on $S^2$  is  expressed in terms  of partial derivatives $\partial_\theta$ and $\partial_\phi$,  while the coefficients depend on $\theta$, only.   Therefore  for fields  which are {\it independent} of ${\alpha}$, i.e. Kaluza-Klein reduced fields, the sum of the two  Laplacians is  the  same as the Laplacians for the four-dimensional subtracted geometry metric, the fact noticed already in \cite{CL11II}.  Note however, that this is the property of Laplacians, only.  Namely, ${\cal R}_{\pm}$, ${\cal L}_{\pm}$ and ${\cal J}_
\pm$, the raising and lowering generators of $SL(2,\R)\times SL(2,\R)\times SO(3)$,  have coefficients in front of derivatives which  explicitly depend on $\alpha$. These generators, when acting on $\alpha$ -independent fields necessarily transform them into $\alpha$-dependent ones, and thus one has to study  representations of fields under the $SL(2,\R)\times SL(2,\R)\times SO(3)$ actions in the full five-dimensional space. In other words, ${\cal R}_{\pm}$, ${\cal L}_{\pm}$ and ${\cal J}_\pm$ 
are not Killing vector fields of the subtracted metric, and thus
do not generate isometries of the subtracted metrics.

\subsection{Truncations of  Five-dimensional Metric}

 We conclude this section with a few remarks on truncations of the five-dimensional metric (\ref{lifted}).  A truncation of this metric to the subspace $d{\tilde \alpha}=0$ results in the four-dimensional metric:
\ben
ds^2_{5\, |{\tilde \alpha}}= \frac{dt^2}{4m^2(\Pi_c-\Pi_s)^2}\, +\frac{dr^2}{X}+d\theta^2+\sin^2\theta\, d\phi^2\, , \label{dat}
\een
which is  a product of the two-dimensional Euclidean space (${\E} ^2$) and a two-shere ($S^2$) with the isometry ${\R}^2\rtimes SO(2)\times SO(3)$.   When setting  $d{\tilde \alpha}=0$, the generators ${\cal L}_i$  turn to those of   $E(2)=\R^2\rtimes SO(2)$,  the  isometries of the two-dimensional Euclidean space $\E ^2$. Note, however, that the truncation to  $d{\tilde \alpha}=0$ slice is not a  Kaluza-Klein reduction  of the five-dimensional space-time\footnote{In the special case with  ${\cal A}=0$,   a reduction on  $d{\tilde \alpha}=0$ results in  the four-dimensional subtracted geometry is  conformal to  $\E ^2\times S^2$ of (\ref{lifted}). These examples have  $a=0$,  $\Pi_s=0$ (see Eq.(\ref{gaugeps}) for the explicit form of ${\cal A}$),  i.e., those are subtracted geometries of static black holes with at least one zero charge and  the conformal factor is $\Delta^{1\over 2}=(2m^3r)^{1\over 2}\Pi_c$.}.

Another slice with $d{\tilde \beta}=0$  reduces the  five-dimensional metric (\ref{lifted}) to $AdS_3\times S^1/Z_2$. The   $SO(2)$ generator is $\partial_\theta$, and  $SL(2,\R)\times SL(2,\R)$ generators  can be expressed in  terms of $r$,$t$, $\phi$ coordinates  which are related to the  global $AdS_3$ coordinates as:
\bea
\sinh^2\rho &= &{r - r_+\over r_+ - r_-}~, \ \ r_{\pm}=m\pm \sqrt{m^2-a^2}\, ,\nonumber \\ 
 \sigma + \tau &= &-{2\pi i \over\beta_H\Omega_H} \phi~, \nonumber \\
  \sigma - \tau &=& - {4\pi i \over\beta_L} ( t - {\beta_R\over 2\beta_H\Omega_H}\phi)~, 
  \label{cooradsp}
\eea
and $\ell_{AdS}=2$.   The above expressions are obtained by substituting   ${\tilde a}=-\phi/[2ma(\Pi_c-\Pi_s)]$, a consequence of $d\tilde \beta=0$, in (\ref{coorads0}) and using Eqs.(\ref{bo}),(\ref{blr}). Note that those are coordinates typically quoted to specify generators $SL(2,\R)\times SL(2,\R)$ in the  four-dimensional theory (see, e.g., \cite{CL97I,CastroFD,CL11II}). However,  this ation of the $SL(2,\R)\times SL(2,\R)$ generators is in our case a consequence of a truncation of the five-dimensional metric to the slice $d{\tilde \beta}=0$ which is not a Kaluza-Klein reduction to the four-dimensional subtracted geometry.

\section{Conclusions}

In this paper we have addressed in further details the origin and properties of subtracted geometries for  general  multi-charged  rotating charged black holes. 
These geometries where originally obtained  \cite{CL11I,CL11II},  by   removing certain terms in the warp factor of the original metric  in such a way that the massless wave equation 
 exhibits the $SL(2,\R)\times SL(2,\R) \times SO(3)$ symmetry. 
We  showed that  these geometries  arise  as a scaling limit of 
multi-charge rotating black holes with three large charges, 
reminiscent of the near-extreme multi-charge rotating
black holes in the dilute gas approximation.  (An analogous scaling limit  for five-dimensional black holes is given in the Appendix.)  
The procedure also allows for the complete determination of sources.  The subtracted geometry depends of on four parameters:  bare mass $m$, bare angular parameter $a$,  and two boost dependent products $\Pi_i\cosh\delta_i$ and $\Pi_i\sinh\delta_i$. Note that the original general black hole solution is determined by six parameters:  bare mass $m$, bare angular parameter  $a$ and four independent boost parameters $\delta_i$.

 We have also shown  that  at least in  the case of the Schwarzschild black hole  the subtracted geometry can be obtained by performing a
, ``infinite boost''  Harrison transformation in  the Einstein-Dilaton-Maxwell gravity on the original unsubtracted 
 black hole.   In the previous version of this paper  it was  conjectured that the subtracted geometry of general multi-charged rotating black holes  arises as a  Harrison transformation of the original 
multi-charged rotating black hole. This has since been confirmed \cite{Vir}.  These results  are significant since they show that the original black hole and  the subtracted geometry clearly lie in the same orbit, specified by a  Harrison transformation. Thus any physical property of the original black hole solution, which is invariant under duality transformation of the theory, remains a property of the subtracted geometry.

In retrospect, since the scaling limit  is closely related to the dilute gas approximation,  this elucidates the  geometry as  near-BPS and its origin  as a Kaluza-Klein 
type quotient   of $AdS_3\times 4 S^2$  with   $SL(2,\R)\times SL(2,\R) \times SO(3)$ symmetry  manifest. 

The subtracted  geometry  is asymptotically conical (AC), and it is reminiscent of the global monopole and  the isothermal gas sphere behavior.    Since the subtraction removes the 
 ambient asymptotically Minkowski spacetime in a 
way that extracts the  $SL(2,\R)\times SL(2,\R) \times SO(3)$ 
symmetry of the black hole, it is  dubbed  ``a black hole in an AC confining  box''. 
Since the subtracted metric has the same  horizon area  and 
periodicity's of the angular and time 
coordinates in the near horizon regions \cite{CL11I,CL11II}  it is  expected to preserve the internal structure of the black hole.
 An important further direction  is a detailed  investigation of its  thermodynamic properties.
  
\vspace{0.5in}
\noindent {\bf Acknowledgments}  \ \ 
We thank Finn Larsen for discussions and collaboration on related topics.  G.W.G thanks the UPenn Department of Physics \& Astronomy for hospitality.
 MC is supported by the DOE  Grant DOE-EY-76-02- 3071, the NSF RTG DMS Grant 0636606, the Fay R. 
and Eugene L. Langberg Endowed Chair  and the Slovenian
Research Agency (ARRS).

\newpage
\section{Appendix: Scaling limit for Five-Dimensional  Black Holes}

In this section, identify the full subtracted geometry of  the general rotating black holes 
in five-dimensional $U(1)^3$ ungauged 
${\cal N}=2$ supergravity. 

    The bosonic sector of the relevant ${\cal N}=2$  five-dimensional theory can be 
derived from the Lagrangian  density:
\ben
e^{-1}\, {\cal L} = R - \frac{1}{2}{\delta\vec\varphi}^2 -
  \frac{1}{4}\sum_{i=1}^3 X_i^{-2}\, {(F^i)}^2  
    + \frac{1}{24} |\epsilon_{ijk}|\, \epsilon^{\mu\nu\rho\sigma\lambda}
  F^i_{\mu\nu}\, F^j_{\rho\sigma}\, A^k_{\lambda}\,,\label{d5lag}
\een
where $\vec\varphi=(\varphi_1,\varphi_2)$, and
\ben
X_1= e^{-\frac{1}{\sqrt6}\varphi_1 -\frac{1}{\sqrt2} \varphi_2}\,,\qquad
X_2= e^{-\frac{1}{\sqrt6}\varphi_1 +\frac{1}{\sqrt2} \varphi_2}\,,\qquad
X_3 = e^{\frac{2}{\sqrt6}\varphi_1}\,.
\een

We write the 5D metric  of the general rotating black hole \footnote{This three-charge rotating black hole is a generating solution for the most general charged rotating black hole  of  maximally supersymmetric give-dimensional ${\cal N}=4$ and ${\cal N}=8$ supergravity theory, which is can be  obtained as a toroidal compactification on an effective heterotic string theory and Type IIA superstring theory, respectively. 
 The most general charged rotating black hole  can be obtained by acting on the generating solution with  a subset of  respective $\{S,T\}$- and $U$- duality transformations.} as a fibration over a 4D base space \cite{CCLPII}\footnote{The base space coordinates $(x,y,\sigma,\chi)$ are 
related to the more familiar radial and angular coordinates  coordinates $(r,\theta,\phi,\psi)$  as
$x = r^2,$ 
$y=a^2 \cos^2\theta + b^2\sin^2\theta~,$ 
 $ \sigma  = {1\over a^2- b^2} \left(a\phi
-b\psi \right)~ $,  $
\chi  = {1\over a^2- b^2} \left( b\phi
-a\psi\right)~.$}

\bea
ds^2_5 & =& - \Delta^{-2/3}_0 G ( dt+{\cal A})^2 + \Delta^{1/3}_0 ds^2_4~,\cr
ds^2_4 &= &{dx^2\over 4X} + {dy^2\over 4Y} + 
{U\over G} (d\chi - {Z\over U} d\sigma)^2 + {XY\over U} d\sigma^2~,
\label{metric5}\eea
where 
\bea
 \Delta_0 =&& (x+y)^3 H_1 H_2 H_3~ , \ \ \ X = (x+a^2) (x+b^2) - 2m x~, \ \ \ 
Y  = - (a^2 - y) (b^2 - y)~, \cr
G = &&(x+y) (x+y - 2m )~, \ \ \  U = yX - xY~, \ \ \ 
Z = ab(X+Y)~, \cr
{\cal A}  = &&{2m \Pi_c \over x + y - 2m}[ (a^2 + b^2 - y)d\sigma - abd\chi]
-{2m \Pi_s \over x+y} (abd\sigma - y d\chi)~,
\eea

The  scalars are given by
\ben
X_i =H_i^{-1}\, (H_1 H_2 H_3)^{1/3}\,,
\een
and gauge potentials by
\bea
A^1=&& \frac{2m}{(x+y)H_1} \{ \sinh{\delta_1}\cosh{\delta_1} \, dt + \sinh{\delta_1} \cosh{\delta_2}\cosh{\delta_3} [ab d\chi 
     + (y-a^2-b^2) d\sigma] \cr +&& \cosh{\delta_1} \sinh{\delta_2} \sinh{\delta_3} (ab d\sigma - y d\chi)\}\, ,
\eea
where  $A^2$and $A^3$ determined by  acting with  cyclic permutations on $\delta_i$ parameters in $A^1$.  Here:
\ben
H_i=1+{{2m\sinh^2\delta_i}\over {x+y}}, \ \ (i=1,2,3),
\een
and we have defined:
\ben
\Pi_c\equiv \prod_{i=1}^3=\cosh\delta_i, \ \ \ \Pi_s\equiv \prod_{i=1}^3
\sinh\delta_i\, .
\een
Note that the solution is parameterized by the bare mass $m$, two rotating parameters $a,b$ and three charge parameters $\delta_i$  $(i=1,2,3)$.

The subtracted geometry for these backgrounds is obtained by taking the scaling limit of the solution where we shall denote all the variables with ``tilde``   and  without loss of generality  taking large two charge parameters equal: ${\tilde \delta}_1=
{\tilde \delta}_2\equiv {\tilde\delta}$. The coordinates and the parameters scale with  $\epsilon\to 0$ as:
\bea
&&{\tilde x}= x \epsilon,   \quad  {\tilde t}=t {\epsilon^{-1}}, \quad  {\tilde y}=y \epsilon, \quad {\tilde \sigma}=\sigma\epsilon^{-1/2},\quad{\tilde \chi}=\chi\epsilon^{-1/2}, \cr
&&{\tilde m}=  m \epsilon, \quad {\tilde a}^2= a^2   \epsilon , \quad  {\tilde b}^2=b^2\epsilon,  \cr
&&2{\tilde m} \sinh^2 {\tilde \delta} \equiv Q = {2m}{\epsilon^{-1/2}} (\Pi_c^2-\Pi_s^2)^{1/2},   \quad \sinh^2{\tilde \delta}_3=\frac {\Pi_s^2}{\Pi_c^2-\Pi_s^2}  \
\eea

The subtracted geometry metric has the same form  (\ref{metric5}) as the general black hole solution except for the subtracted warp factor:
\ben
\Delta_0\to \Delta = (2m)^2 
(x+y) (\Pi^2_c -\Pi_s^2) +(2m)^3\Pi_s^2 ~.
\een
This geometry is sourced by the scalar fields:
\ben
X_1=X_2=X_3^{-\frac{1}{2}}=\frac{\Delta^{\frac{1}{3}}}{2m}\, ,
\een
and the gauge potentials:
\bea
&&A^1=A^2=-\frac{x+y}{2m}\, dt+y\Pi_c\,  d\sigma-y\Pi_s \, d\chi \, , \cr
&&A^3=\frac{(2m)^4\Pi_s\Pi_c}{(\Pi_c^2-\Pi_s^2)\Delta}\, dt+\frac{\Pi_s}{\Delta}[ab\, d\chi +(y-a^2-b^2)d\sigma]+\frac{\Pi_c}{\Delta}(ab\, d\sigma-y\, d\chi)\, .
\eea
Note that we  have chosen a gauge where we  have rescaled the scalars  and the field strengths by appropriate factors  of $\epsilon$ and $\Pi_c^2-\Pi_s^2$.
The solution is of co-homogeneity two, with gauge field strengths having both electric and magnetic components.

The scaling limit,  reminiscent of the dilute gas approximation,  extracts the subtracted geometry  of the five-dimensional black hole which  is a Kaluza-Klein coset of $AdS_3\times S^3$ exhibiting conformal invariance. It is a solution of the six-dimensional Lagrangian (\ref{d6lag}) with $F^3$ corresponding to  the Kaluza-Klein field strength. The scaling limit also signifies that  the geometry is supersymmetric.

\end{document}